\documentclass[preprint2]{aastex}
\usepackage{graphicx}

\begin{document}

\title{Nothing to hide: An X-ray survey for young stellar objects in the Pipe Nebula}

\author{Jan Forbrich, Bettina Posselt, Kevin R. Covey\altaffilmark{1}, \& Charles J. Lada}
\affil{Harvard-Smithsonian Center for Astrophysics, 60 Garden Street, Cambridge, MA 02138, USA}
\altaffiltext{1}{now at: Department of Astronomy, Cornell University, Ithaca, NY 14853, USA}

\begin{abstract}
We have previously analyzed sensitive mid-infrared observations to
establish that the Pipe Nebula has a very low star-formation
efficiency.  That study focused on YSOs with excess infrared emission
(i.e, protostars and pre-main sequence stars with disks), however, and
could have missed a population of more evolved pre-main sequence stars
or Class III objects (i.e., young stars with dissipated disks that no
longer show excess infrared emission).  Evolved pre-main sequence
stars are X-ray bright, so we have used ROSAT All-Sky Survey data to
search for diskless pre-main sequence stars throughout the Pipe
Nebula.  We have also analyzed archival XMM-\textsl{Newton} observations of three
prominent areas within the Pipe: Barnard 59, containing a known cluster of young
stellar objects; Barnard 68, a dense core that has yet to form stars;
and the Pipe molecular ring, a high-extinction region in the bowl of
the Pipe. We additionally characterize the X-ray properties of YSOs in Barnard 59.  
The ROSAT and XMM-\textsl{Newton} data provide no indication of a
significant population of more evolved pre-main sequence stars within
the Pipe, reinforcing our previous measurement of the Pipe's very low
star formation efficiency.
\end{abstract}


\section{Introduction}

In a study using deep mid-infrared imaging, we found a very low level of star formation activity in the Pipe Nebula ($10^4$~$M_\odot$, 130~pc, \citealp{lom06}), with an upper limit of only 21 candidate YSOs in the entire cloud \citep{for09}. This survey was most sensitive to young stellar objects (YSOs) with infrared excesses and presumably in the earliest stages of evolution. A population of more evolved class III YSOs, i.e., diskless young stellar objects without infrared excess (e.g., \citealp{lad87}), if present, would have important implications for the cloud. Most importantly, such a population would lead to an increased estimate of the region's star-formation efficiency and would also indicate that the molecular cloud is more long-lived than otherwise thought.

Class III objects can be identified by their X-ray emission. While this technique has been known for some time (e.g., \citealp{kuc79}), it has been strengthened by the analysis of the Chandra Orion Ultra-deep Project which produced a statistically well defined description of the X-ray properties of YSOs \citep{pre05}. Furthermore, class III objects have been studied in \textsl{Spitzer} infrared surveys of other nearby ($d\sim$120--160~pc) star-forming regions. \citet{pad06} targeted 83 class III sources in the Taurus, Lupus, Ophiuchus, and Chamaeleon star-forming regions at 24~$\mu$m, detecting all 83 but identifying excesses in only 5. \citet{cie07} also used c2d data to study 102 spectroscopically identified class III sources in the Lupus and Ophiuchus regions\footnote{leaving out the more distant IC~348 in Perseus for the sake of this comparison}, detecting 80\% of them at 24~$\mu$m (82/102) at a sensitivity comparable to our data and with distances of about that of the Pipe. Requiring mid-infrared (MIPS) as well as X-ray detections thus provides better constraints on class~III source populations than any of these two datasets alone could provide, and we use this additional constraint in our analysis of the Pipe Nebula.

In this study, we first use the ROSAT All-Sky Survey (RASS) as the only available means to search the entire Pipe Nebula complex for X-ray--active stars. We then study three regions of particular interest using higher-sensitivity data obtained by the X-ray satellite XMM-\textsl{Newton}. The latter satellite has considerably more collecting area and is sensitive to harder X-rays than ROSAT, allowing observations of more extincted sources. We use XMM-\textsl{Newton} to probe three prominent regions in the Pipe Nebula complex in more detail; Barnard 68 (B\,68), Barnard 59 (B\,59), and the Pipe molecular ring in the bowl of the Pipe \citep{mue07}, in the following referred to as the PMR. While B\,59 is known to contain a cluster of YSOs \citep{bro07,cov10}, the entire bowl area has been found to be devoid of protostars \citep{for09}, even though the average extinction in that area is at levels of visual extinction higher than those of B\,59. 

The Pipe region is seen in projection against the Galactic bulge, providing an enormous background source density, particularly at near-infrared wavelengths. Therefore, we have to compare the density of X-ray sources inside and outside of the identified dense core population (from \citealp{rat09}) to estimate the local density of background X-ray sources. In the following, we use the same extinction criterion in defining the `on-core' regions that was also used by \citet{rat09}: These regions have a minimum optical extinction, $A_V$(sub)$>1.2$~mag, in an extinction map in which a slowly varying background has been removed \citep{alv07}. We have used the same criterion in our \textsl{Spitzer} study \citep{for09}. For purposes other than defining the on-core regions, we use the total extinction map computed by \citet{lom06}. The pixel size of the extinction maps is $30''$. 

Looking for possible class~III objects, we select X-ray sources with 2MASS counterparts in these regions, requiring $K_S$ band detections. We then use their X-ray properties as well as \textsl{Spitzer} IRAC and MIPS data to better characterize them. Finally, to further constrain any population of class~III sources, we combine the X-ray source selection with detections at 24~$\mu$m from our sensitive \textsl{Spitzer}-MIPS study \citep{for09}.

\section{Observations}

To cover the entire Pipe Nebula complex, we make use of the ROSAT All-Sky Survey in the form of the Bright Source Catalogue \citep{vog99} and the Faint Source Catalogue \citep{vog00}. The observations for this survey were carried out in 1990/91 with the Position-Sensitive Proportional Counter (PSPC), in the energy range of 0.1--2.4~keV.

The X-ray satellite XMM-\textsl{Newton} has observed the Pipe Nebula region several times. Here, we analyze three particular observations, targeting the prominent extinction cores B\,59, FeSt 1-457\footnote{Note that although this observation is listed as targeting FeSt 1--457, it is pointed at the neighboring PMR.}, and B\,68. The latter region was observed for 51~ksec on September 27, 2002 (07:19--21:00~UT, median filter, observation ID 0152750101), followed by observations of the two other regions on August 29, 2004: Fest 1-457 (06:19--19:53~UT, 53~ksec, thin filter, observation ID 0206610201) and B\,59 (21:53--10:14(+1)~UT, 49~ksec, thin filter, observation ID 0206610101). All of these observations have been processed as part of the 2XMMi catalog \citep{wat09}. Note that pipeline products like spectra and light curves for all catalog sources can be downloaded from a dedicated website\footnote{e.g., \url{http://xcatdb.u-strasbg.fr/2xmmi}}. The good time intervals for these observations amount to observing times of 48.2~ksec for B\,59, 43.8~ksec for the PMR, and 44.8~ksec for B\,68. The energy range accessible to XMM-\textsl{Newton} and used for the 2XMMi catalog is wider than the one of ROSAT; it spans 0.2--12~keV. In the following, we mainly work with this catalog, but we have also reprocessed the data using SAS 8.0.0 to subsequently run our own source detection using the SAS task ewavelet. We also employ the SAS task especget to produce X-ray spectra of selected sources. For this purpose, we mostly use the EPIC-pn data only, rather than EPIC-MOS data, apart from the unresolved sources [BHB2007] 1\&2  \citep{bro07} which are so far off-axis ($r=11\farcm9$) that they are only detected with the EPIC-MOS detectors. Typically, spectra were extracted from areas with a radius of 25$''$ and similarly sized, nearby background regions. The only exception is the close pair [BHB 2007] 6 \& 7, where we chose a radius of 10$''$ in an attempt to separate the two components. The task especget produces the corresponding ancillary response and response matrix files. Spectra were finally binned with minimum bin sizes of 20 counts. Fits were carried out using the CIAO tool \textsl{Sherpa}, V3.4 \citep{fre01}.

In order to assess the infrared spectral energy distributions of selected sources, we have performed aperture photometry on \textsl{Spitzer}-IRAC data covering the region, most notably observation IDs 5132288, 14089472, and 14090496. For this purpose, we have used pipeline PBCD data from the \textsl{Spitzer} archive. The typical aperture radius was 3 image pixels with an annulus for background subtraction with inner and outer radii of 10 and 20 pixels. We have additionally used \textsl{Spitzer}-MIPS data as discussed in \citet{for09}.

\section{Analysis}

\subsection{The ROSAT All-Sky Survey}
\label{sec_rosat}

\begin{figure*}
\centering
      \includegraphics*[width=\linewidth]{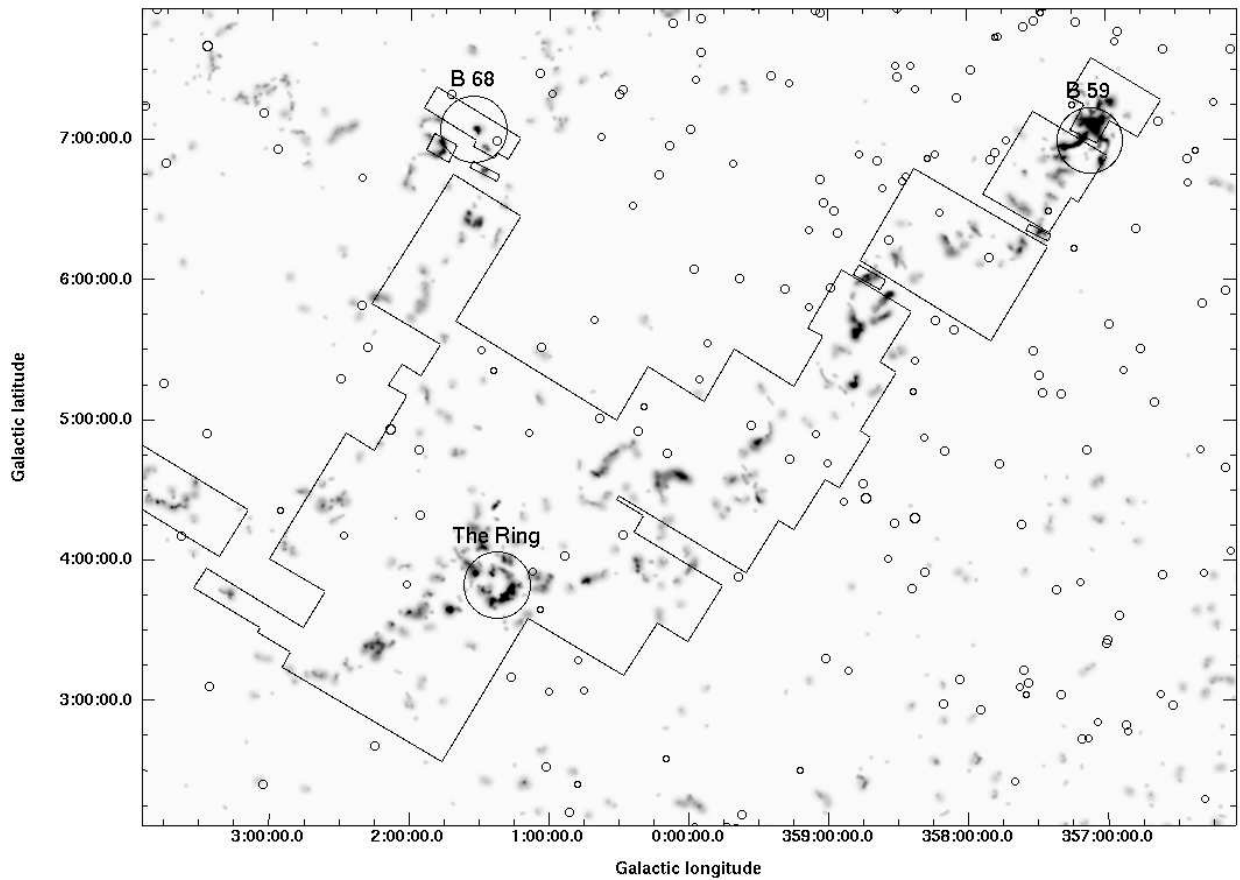}
\caption{A background-subtracted extinction map \citep{alv07} of the Pipe Nebula with ROSAT All-Sky Survey sources shown as small circles. Three large circles mark the areas studied in more detail using XMM-\textsl{Newton}, indicating the usable fields of view. The continuous line delineates the area surveyed with \textsl{Spitzer}-MIPS \citep{for09}. \label{fig_rosat}}
\end{figure*}

The area covered in our mid-infrared survey \citep[][Fig.~8]{for09} contains 44 ROSAT All-Sky Survey sources, i.e., 12 and 32 sources from the Bright and Faint Source Catalogues, respectively. The locations of these sources are shown in Figure~\ref{fig_rosat}, overlaid on an extinction map of the Pipe Nebula. Given typical $1\sigma$ positional uncertainties of $\sim20''$, it is nearly impossible to reliably identify near-infrared or optical counterparts to ROSAT X-ray sources in this region with its high density of background sources, simply due to confusion. While this is somewhat easier at mid-infrared wavelengths due to the lower source density in this wavelength range, all 36 sources whose 3$\sigma$ error circles are located entirely within the MIPS source detection coverage have one or more mid-infrared counterparts at 24~$\mu$m. Eight additional sources have error circles that only partly overlap with the area where MIPS source detection was performed. Note that the error circle of only one ROSAT source (1RXS J173319.3--255416) overlaps with the on-core regions and all but two of the sources have zero extinction in the extinction map where the extended structure has been removed (see above). Their median total extinction is $A_V=2.4$~mag. Correlating the source list with catalogs using SIMBAD shows that two of these ROSAT sources have been identified as X-ray binaries, two as Hipparcos stars (at 54 and 149~pc, respectively) and one as an RS~CVn binary.

To assess the completeness of the ROSAT All-Sky Survey for our purposes, we determine the limiting unabsorbed X-ray luminosities for different levels of foreground extinctions. The limit of six source counts in the Faint Source Catalog corresponds to a limiting count rate of 0.02\,s$^{-1}$ with the typical exposure time of 300\,s in this area. We can now simulate X-ray sources with a given spectrum to find the limiting unabsorbed X-ray fluxes and in turn the luminosities corresponding to this count rate limit. Using the Portable, Interactive Multi-Mission Simulator \citep[PIMMS\footnote{\url{http://heasarc.nasa.gov/docs/software/tools/pimms.html}}, ][]{muk93}, we determine such unabsorbed X-ray luminosities for sources using simulated Raymond-Smith thermal plasma spectra with temperatures of $\sim10$~MK. While the median total visual extinction from our extinction map for the 44 ROSAT sources is $A_V=$2.4\,mag, we calculate the limiting luminosities also for $A_V=$0,5, and 25~\,mag, the maximum of our total extinction map (which constitutes only a lower limit for the actual maximum extinction in the cluster). The visual extinctions can be converted into absorbing hydrogen column densities using the empirical relation $N_H$[cm$^{-2}$]$\approx2\times10^{21}\times A_V$[mag] \citep[e.g., ][]{ryt96,vuo03}. Conservatively assuming a plasma temperature of 1~keV \citep[e.g.,][]{fei05}, the limiting luminosities are log($L_X$)=29.5, 30.2, 30.7, and 32.3\,erg\,s$^{-1}$ for the given visual extinctions (in rising order), assuming a distance of $d=130$\,pc. Note that the high-extinction regions cover only a minute fraction of the complex.

We can now make use of the cumulative X-ray luminosity distribution functions derived for the PMS population of the Orion Nebula Cluster \citep{pre05}, to assess the completeness of the ROSAT data for G, K, and M pre--main-sequence stars. The limiting unabsorbed X-ray luminosity that we have determined for the median foreground extinction of $A_V=2.4$\,mag corresponds to a completeness of $\sim55$\% for G and K stars and $\sim15$\% for M stars. For $A_V=5$\,mag, the completeness diminishes to $\sim30$\% for G and K stars and close to zero for M stars while without foreground extinction, the numbers are $\sim90$\% completeness for G and K stars and $\sim50$\% for M stars.

A significant subset of the selected ROSAT sources can be expected to be located in the Galactic bulge against which the Pipe Nebula is projected, to be of extragalactic nature, or to be in the foreground. To assess this aspect, we determine the source density in the Pipe and in neighboring regions (shown in Fig.~\ref{fig_rosat}). The overall RASS source density in the region covered in our mid-infrared survey, where we find 44 RASS sources, is 3.29 sources per square degree. However, it turns out that the RASS source density in the Pipe Nebula is \textsl{lower} than in the immediately surrounding regions. 
Our Pipe Nebula total extinction map covers nearly 49 square degrees which is about 3.7 times more than the area covered in the mid-infrared survey (see Figure~8 in \citealp{for09}). This area contains a total of 199 RASS sources. In the immediate surroundings of the high-extinction regions inside the infrared survey area, we thus find a RASS source density of 4.37 sources per square degree. This source density is a good measure of the contaminant source density in this region since the median total extinction of the RASS sources inside the infrared survey area ($A_V=2.4$~mag) and the median extinction for RASS sources in the remainder of the area covered by the extinction map ($A_V=2.0$~mag) are very similar. Extrapolating from the source density in the surroundings, we would expect to find 58.4 sources in the infrared survey area, but we only find 44 of which six are known to be unrelated to the Pipe Nebula. The Poisson error ranges of these two numbers do not overlap. It appears likely that the source density in the Pipe Nebula is lowered by the partial extinction of a distribution of background sources, for example in the Galactic bulge or extragalactic sources, by the molecular cloud. This effect is difficult to quantify since it depends on the intrinsic source spectra, e.g., on whether a given class of sources is rather soft and therefore more susceptible to extinction. However, while in our total-extinction map of the Pipe Nebula and its surroundings, 11.5\% of the pixels are of extinctions $A_V>5$~mag, only 6.8\% of the RASS sources in the same area are affected by the same amount of extinction, as derived from the same extinction map, consistent with a picture in which we do not detect the most heavily extincted sources.

Without sufficient information about optical or infrared counterparts, we can only make use of the X-ray source properties for further characterization. For the ROSAT sources, leaving aside the potentially variable count rates, these are primarily the two standardized hardness ratios HR1 and HR2 \citep{vog99}, ranging from --1 to +1. Similar to, e.g., infrared colors, the hardness ratios provide a crude characterization of spectral properties in the absence of actual spectra. The second ratio probes harder X-ray emission within the ROSAT band than the first hardness ratio. However, among the 44 ROSAT sources selected here only six have hardness ratios HR1 and HR2 with errors of $<0.3$.

In order to judge even these few hardness ratios, we need a reference region. Since the Taurus star-forming region is of about the same size, mass and distance as the Pipe Nebula, we compare our sample with a correlation of Taurus PMS stars and the ROSAT All-Sky Survey, using the listing of all known Taurus PMS stars by \citet{ken08}.

\begin{figure*}[h]
\centering
 \includegraphics*[width=0.5\linewidth]{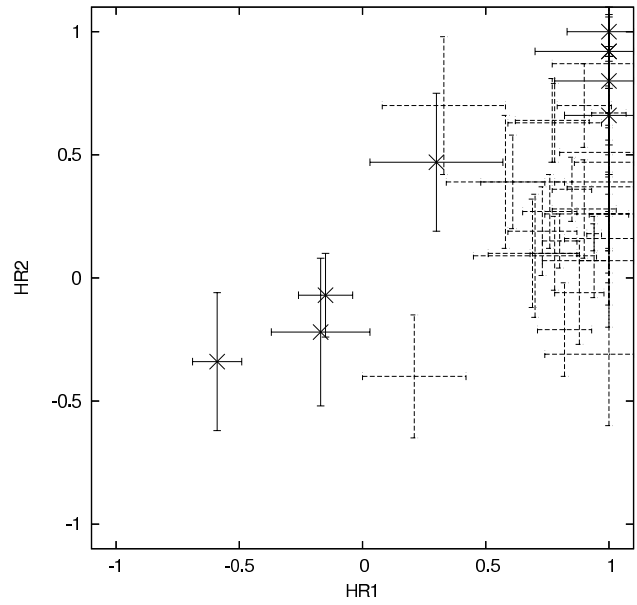}
\caption{ROSAT hardness diagram (HR1 and HR2) for the sources in the Pipe Nebula region, `$\times$' with solid lines, and Taurus PMS sources from \citet{ken08}, dashed lines. Data only shown for sources that have both hardness ratios, HR1 and HR2, with errors $<0.3$. \label{fig_rosat_2}}
\end{figure*}

While this list is not complete, it provides us with a representative comparison sample. Conservatively using 2$\sigma$ error circles\footnote{Using 1$\sigma$ errors would yield 24 sources, while a 3$\sigma$ search radius yields 56 sources. The median 2$\sigma$ search radius is $28''$.}, there are 46 RASS sources encompassing at least one PMS object from this Taurus list. This overall number of sources is similar to the results of the ROSAT Taurus survey by \citet{neu95}. Out of these 46 sources, 29 have hardness ratios with reasonable errors $<0.3$. In Fig.~\ref{fig_rosat_2}, we show a plot of the two hardness ratios for the Pipe and the Taurus samples. While the latter sources delineate the same region of the diagram that \citet{neu95} found to be occupied by class~II and III objects, only about half of the Pipe sources are in that region, although the overall numbers are very low. Unfortunately, PMS objects are not the only source class with predominantly positive hardness ratios: X-ray binaries fall onto the same area of the hardness ratio diagram (e.g., \citealp{hap99,hab00}). To show only one example: While both the only ROSAT source in the on-core region and the brightest ROSAT source in our sample have HR1=HR2=1, the latter is a previously known X-ray binary that is unrelated to the Pipe Nebula. As a further complication, the intrinsic hardness ratios of background sources are affected by foreground extinction. The hardness ratios therefore do not provide significant further constraints for our purposes. 

In summary, the ROSAT source density in our Pipe Nebula mid-infrared survey area is lower than in the immediate surroundings; 39 of 44 sources remain unidentified. The on-core regions contain at most a single RASS X-ray source. Of all 44 sources, their median total extinction is $A_V=$2.4\,mag, only six have hardness ratios of sufficient quality to compare them to Taurus PMS objects, and only about half of them fall onto the same region of the hardness diagram as the Taurus sources, even though that parameter space of hardness ratios is not unique to PMS sources. 
We conclude that likely \textsl{none} of the RASS sources located toward the Pipe Nebula are class III YSOs. 
To substantiate this result, we take a deeper X-ray look at three high-extinction regions in the Pipe Nebula using XMM-\textsl{Newton}.

\subsection{XMM-\textsl{Newton} data}

As noted above, all of the XMM-\textsl{Newton} observations considered here have been processed as part of the 2XMMi catalog \citep{wat09}. We use this catalog to search for X-ray sources with 2MASS-$K_S$ band counterparts, particularly in the on-core regions. Ancillary \textsl{Spitzer} MIPS and IRAC data allow us to subsequently study the SEDs of identified sources in more detail. Only one ROSAT source is located within the fields of view of the three XMM-\textsl{Newton} observations (1RXS~J172236.0--235938 near the southern edge of the B\,68 field; it is compatible with the position of a 2XMMi source).

\subsubsection{Initial X-ray source selection}

In a first step, we exclude low-quality detections from the X-ray sources listed in the 2XMMi catalog for the three different observations. After a visual check, we exclude five sources listed as likely spurious\footnote{We exclude sources with a 2XMMi summary flag of 4.}. Out of 297 initially listed sources, this leaves 292. Only a minor filtering of detection likelihoods has been applied; we conservatively use a minimum detection likelihood of srcML=8, as provided by the 2XMMi catalog (where the minimum detection likelihood reported is srcML=6). 
We exclude 17 sources that were detected as extended objects, identified by their catalogued extension likelihood. These two steps of filtering leave 251 X-ray sources. The distribution among the three observations is as follows: 112 of these sources are located in the field of view of the B\,68 observation, 101 in the field of B\,59, and 38 in the PMR.  

Any list of X-ray detections will likely include background sources in the Galactic bulge, and we note that the three fields covered by \textsl{XMM}-Newton cover a total of 43704 2MASS sources. We therefore again separate parts of the analysis into on-core and off-core regions. It is important to keep in mind that the on-core regions cover different fractions of the three fields of view. While these fractions are 49\% and 39\% for B\,59 and the PMR, respectively, only 14\% of the B\,68 field of view consists of on-core regions. The average total extinction in the three fields of view, as measured in our total extinction map, is highest in the PMR ($A_V=8.7\pm2.6$~mag), followed by a similarly extincted B\,59 ($A_V=5.4\pm3.5$~mag) and a quite different B\,68 area at only $A_V=1.2\pm1.2$~mag. In contrast to the other two regions, there is virtually no extinction surrounding the B\,68 core. These basic properties of the three fields as well as results of the subsequent analysis are summarized in Table~\ref{tab_2xmmi_sel}.

Among the X-ray properties, we only make use of the X-ray count rates in the following. The significant extinction influences the intrinsic hardness ratios of background sources so that hardness ratios from reference regions cannot be directly applied. As we have seen in the discussion of ROSAT data, it is nearly impossible to identify PMS objects based on their X-ray hardness ratios alone. Therefore, the X-ray hardness ratios provided by the 2XMMi catalog do not allow us to constrain source classes.

\subsubsection{Infrared counterparts and SED analysis}

In a second step in our search for candidate class~III sources, we only select X-ray sources with 2MASS counterparts and, more specifically, $K_S$-band detections. To define a cutoff $K_S$ band magnitude, we use the fact that 90\% of the unique 2MASS counterparts to the Taurus PMS sources listed by \citet{ken08} have $K_S<12.4$~mag, using all cases where the source is within the 3$\sigma$ 2MASS error circle. While the source density in the Pipe Nebula region is very high due to its projection against the Galactic bulge, we can thereby at least eliminate some X-ray sources without sufficiently bright NIR counterparts. Note that the total number of 2MASS sources fulfilling this $K_S$ band cutoff criterion is still 7630 for the three pointings. 

For every 2XMMi source selected thus far, we have extracted the X-ray positions corresponding to the three observations that we consider and we look for 2MASS counterparts in the 3$\sigma$ error circles as quoted by the 2XMMi catalog\footnote{The 2XMMi catalog uses a systematic error of either $0\farcs35$ or $1\farcs00$, depending on whether an acceptable astrometric correction was obtained. Among the datasets considered here, the smaller error applies only to the B\,59 observation. The infrared counterparts of two known YSOs in B\,59, imaged as bright sources at large off-axis angles ($r=11\farcm9$) and including one potential unresolved double X-ray source, lie just outside the total 3$\sigma$ error circles as listed in the 2XMMi catalog. We therefore assume a systematic error of $0\farcs50$ for B\,59 instead of $0\farcs35$.}. The radius of the median 3$\sigma$ error circle is $3\farcs9$ where the values range between $1\farcs6$ and $8\farcs4$. Starting with the list of 251 X-ray sources, we thus find 46 X-ray sources with 2MASS counterparts of $K_S<12.4$~mag. One of the sources has two suitable 2MASS counterparts in the search radius in which case we choose the one closer to its X-ray position. Note that the positions of 10 out of these 46 X-ray sources are also compatible with weaker 2MASS sources.

Given the high 2MASS source density in the target regions, we can now assess the possibility of chance alignments, i.e., the number of 2MASS sources above our $K_S$ band cutoff that we would find if both the X-ray search areas and the 2MASS sources were randomly distributed across the fields of view. A summarized in Table~\ref{tab_2xmmi_sel}, we would globally expect about 17 of the 46 selected sources to be chance alignments. Not differentiating between on- and off-core regions, only the number of selected sources in the B\,59 and B\,68 fields is significantly above the expected number of such chance alignments.

To find out whether any of the 46 selected X-ray sources with near-infrared counterparts, including the subgroup of 13 sources in the on-core regions, have spectral energy distributions indicative of their YSO status, we have obtained \textsl{Spitzer} mid-infrared aperture photometry for these sources, using the four IRAC bands in addition to our previously discussed MIPS 24\,$\mu$m data. While any candidate class III sources might simply show photospheric SEDs, we here primarily look for previously missed excess sources. All selected X-ray sources in on-core regions and most in the off-core regions were covered in all IRAC bands and at 24~$\mu$m and 70~$\mu$m (MIPS). 
Apart from the known YSOs, all X-ray sources with NIR counterparts and \textsl{Spitzer} photometry were subsequently fitted using the online SED fitter by \citet{rob07} to check both YSO models and extincted stellar photosphere fits.  The model fits reveal that all but one of these sources have SEDs that can be reasonably explained by stellar photospheres\footnote{The exception is 2XMMi J173400.2-253746 where only the 24\,$\mu$m flux is significantly brighter than photospheric (the source is undetected at 70~$\mu$m). This source is outside of, but close to the PMR on-core region.}.
The extinction derived from these SED fits is, in most cases, lower than the estimate based on the extinction map by 1--2~mag, indicating that the objects are located inside the cloud. In the case of source 5 (2XMMi~J171117.5--272535), the SED fit requires an extinction of only $A_V=0.3$~mag while the corresponding pixel of the extinction map indicates $A_V=13.0$~mag. A fit to the X-ray spectrum supports a low absorbing column density\footnote{While an absorbed APEC fit to 101 total EPIC-pn counts does not produce statistically meaningful parameter estimates, it yields a 3$\sigma$ upper limit for the extinction of $A_V<0.7$~mag, using the above-mentioned conversion.}, as do the soft hardness ratios of this source, as catalogued. This source, therefore, probably is a foreground star.

\subsubsection{Final source selection and assessment}

To better constrain potential candidate YSOs, we can apply a lower limit to the X-ray fluxes, knowing that X-ray counterparts to any YSOs would be relatively bright X-ray sources, at least in extinction-free regions. We thus exclude all sources with unabsorbed X-ray fluxes that are lower than those derived from the X-ray luminosities of 90\% of the Orion PMS M-stars \citep{pre05}, i.e., $F_X<2.5\times10^{-14}$ erg\,s$^{-1}$\,cm$^{-2}$. In order to determine the unabsorbed X-ray flux for a given source, we use PIMMS to convert the total count rate (0.2--12\,keV), conservatively assuming that the spectrum is thermal plasma emission that can be explained by a 1~keV APEC model and the extinction from the total extinction map. Using this criterion, we can exclude nine low-extinction (and off-core) objects of the 14 sources in the B\,68 field, reducing the sample to a total of 37 X-ray sources with 2MASS counterparts. The X-ray sources in the surroundings of the B\,68 core appear to be unrelated to the cloud.

\begin{figure*}
\centering
      \includegraphics*[width=\linewidth]{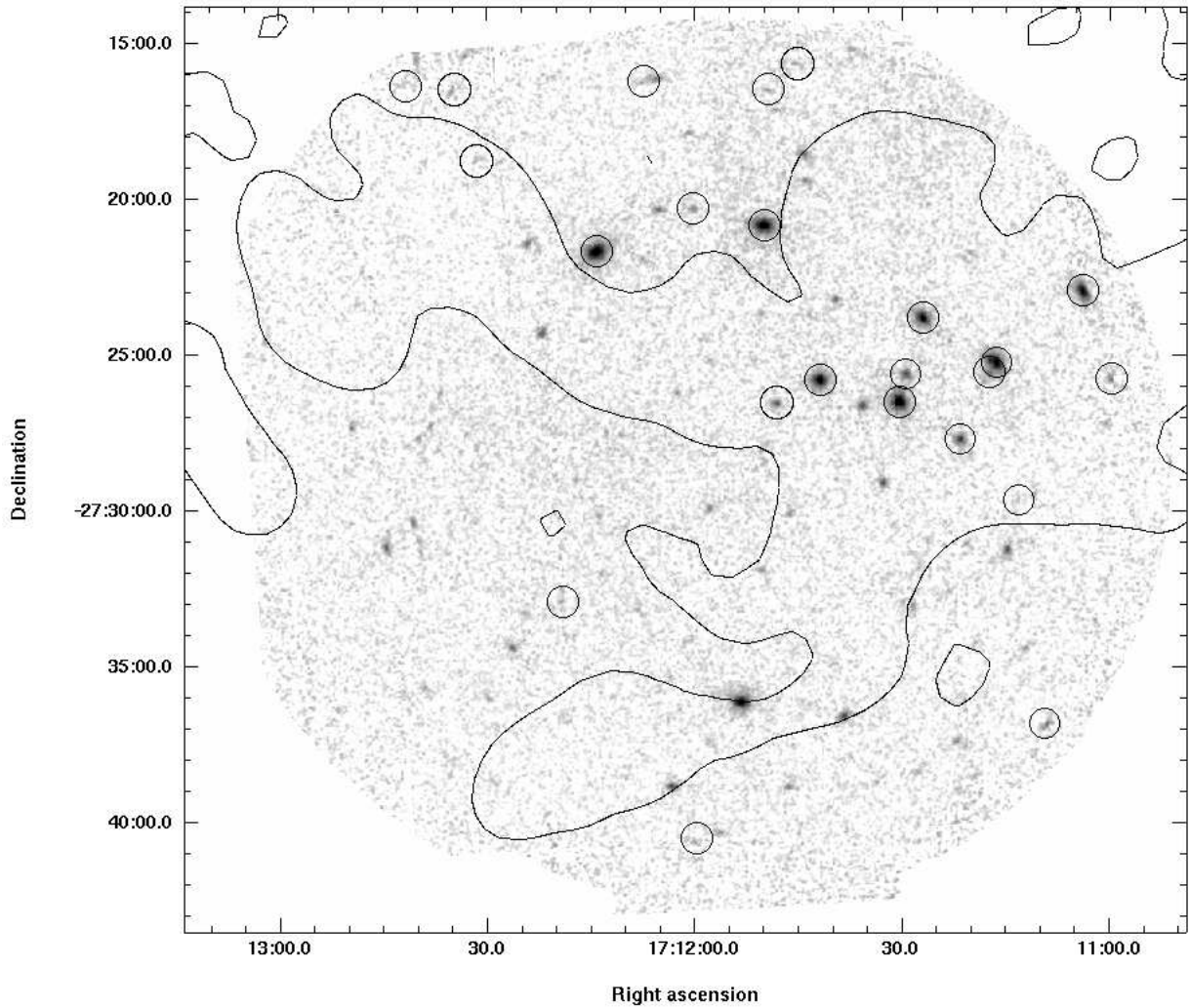}
\caption{XMM-\textsl{Newton} view of B\,59, using merged data of the PN and MOS1/2 detectors. Contour lines show the $A_V=1.2$\,mag levels from the background-subtracted extinction map, delineating the ``on-core'' regions. Circles show 2XMMi sources with 2MASS counterparts having $K_S$-band detections ($<12.4$~mag).\label{fig_xmm_b59}}
\end{figure*}

\begin{figure*}
\centering
      \includegraphics*[width=\linewidth]{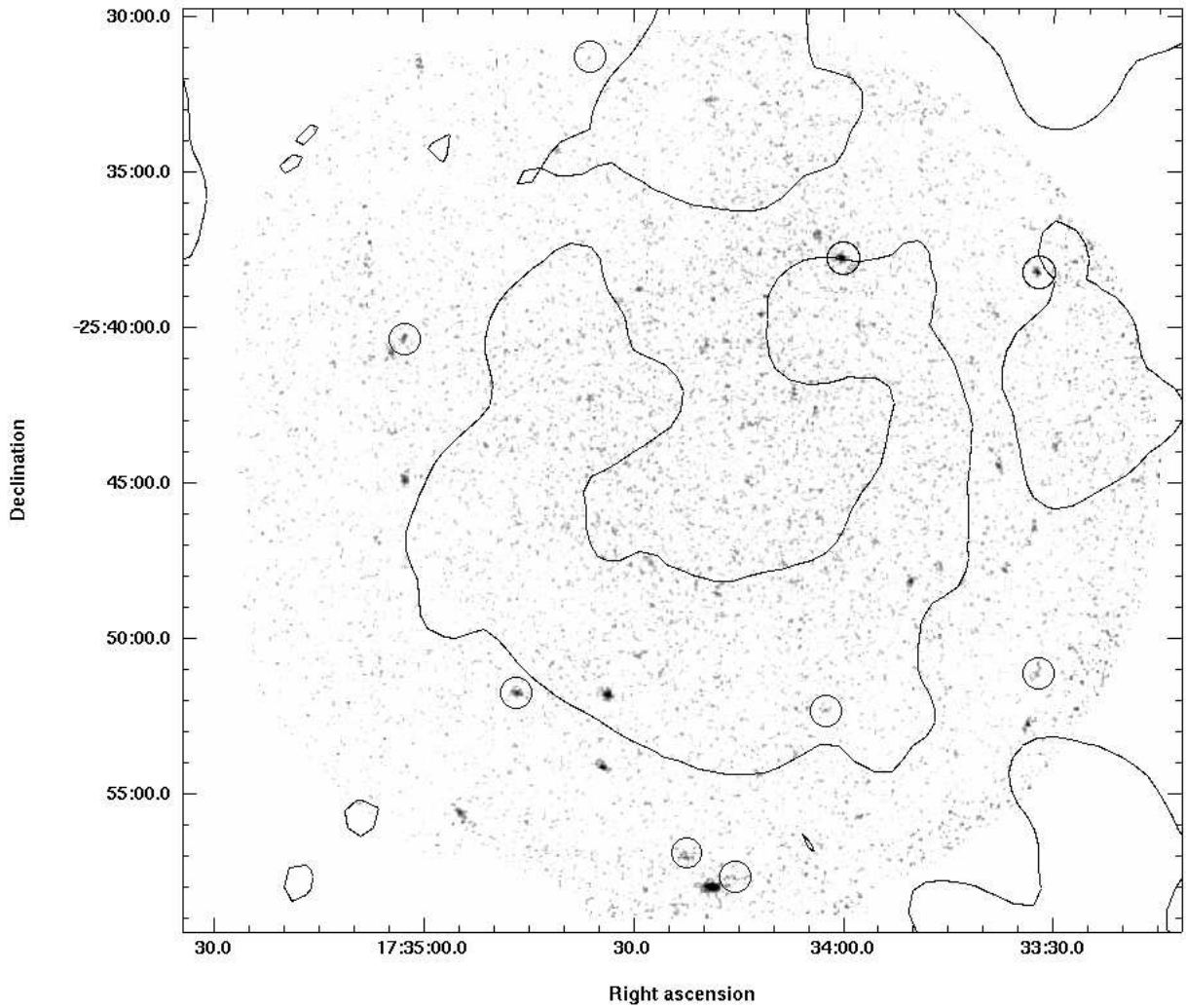}
\noindent
\caption{XMM-\textsl{Newton} views of the Pipe molecular ring, using merged data of the PN and MOS1/2 detectors. Contour lines and symbols as in Fig.~\ref{fig_xmm_b59}.\label{fig_xmm_pmr}}
\end{figure*}

\begin{figure*}
\centering
      \includegraphics*[width=\linewidth]{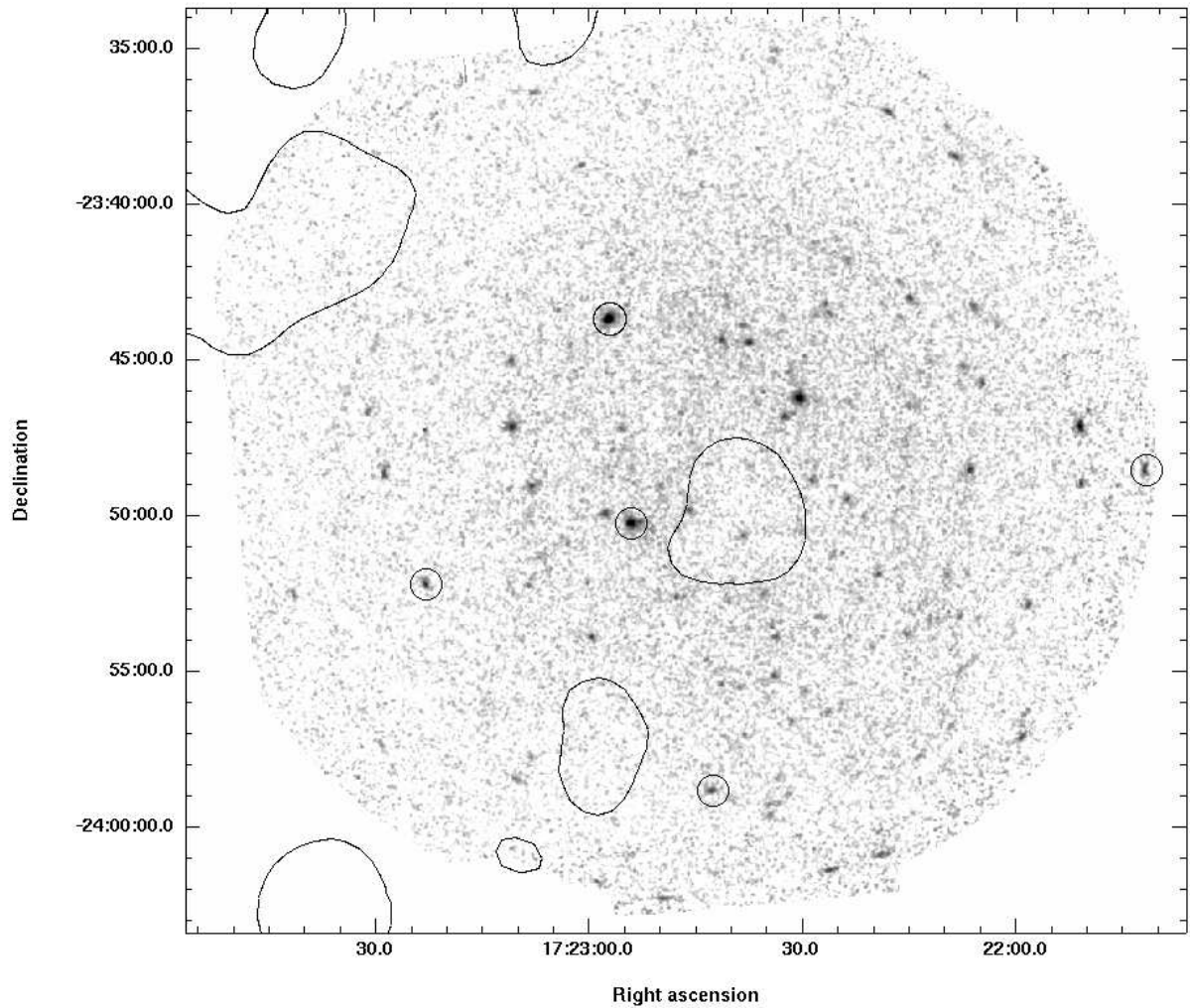}
\noindent
\caption{XMM-\textsl{Newton} views of B\,68, using merged data of the PN and MOS1/2 detectors. Contour lines and symbols as in Fig.~\ref{fig_xmm_b59}. Note that only the final selection of sources is shown. Among the three observations, this field has by far the lowest extinction in the off-core regions (see Table~\ref{tab_2xmmi_sel}).\label{fig_xmm_b68}}
\end{figure*}

Of these 37 sources, 23 sources are located in the B\,59 field, 9 in the PMR, and 5 in the B\,68 field. A view of the three X-ray datasets and the selected sources is shown in Figures~\ref{fig_xmm_b59}, \ref{fig_xmm_pmr}, and \ref{fig_xmm_b68}. Among these are 13 sources in the on-core regions of B\,59 (12) and the PMR (1), with none in the on-core regions of the B\,68 field. These sources are listed in Table~\ref{tab_2xmmi}. The single source in the PMR and one source in B\,59 are close to the $K_S$-band cutoff. Only the number of on-core sources in the B\,59 fields surpasses the number of expected chance alignments, as summarized in Table~\ref{tab_2xmmi_sel}. The difference between B\,59 and the PMR is striking, given that both are comparable in overall extinction and amount of on-core regions covered.

Among the 13 X-ray sources with 2MASS-$K_S$ counterparts in the on-core regions of B\,59, there are 7 previously identified YSOs from \citet{bro07}, not differentiating [BHB2007] 1 and 2 which are unresolved in the XMM-\textsl{Newton} data due to the large off-axis angle. Additionally, [BHB2007] 7 appears to have a weak X-ray counterpart that is not listed in the 2XMMi catalog, but was identified in our wavelet source detection. One X-ray detected YSO -- [BHB2007] 20 -- is located outside of the on-core region, and the deeply embedded protostar [BHB2007] 11, while X-ray--detected, is filtered out\footnote{Note that in our search for class~III sources, the criteria were not designed to include protostars.} due to its weak 2MASS counterpart. In the innermost, most deeply embedded part of the \textsl{Spitzer}-identified young cluster \citep{bro07}, no new candidate members were detected. However, a 2MASS source south of [BHB2007] 6 and 7 (not among the identified YSOs) was found to have a weak, soft X-ray counterpart. The X-ray properties of the known YSOs in B\,59 are discussed in more detail in Section~\ref{sec_b59}.

Since we undoubtedly again detect background sources in the Galactic bulge next to putative sources in the cloud as well as foreground sources, we again analyze the density of X-ray sources in the on-core and the off-core regions. In B\,68, all X-ray sources with NIR counterparts lie in off-core regions; this is not the case in B\,59 and the PMR. In the absence of a class~III source population in the Pipe region, we would expect to find a source density that is lower than or equal to the source density in the off-core region due to the additional extinction. In B\,59, with its known cluster of YSOs, we find that the source density in the on-core regions is indeed slightly higher than the source density in the surrounding off-core regions (52\% of the X-ray sources with NIR counterparts lie in the on-core regions which cover 49\% of the field of view). In the PMR, in striking contrast, we find that only 11\% of these sources are in the on-core regions even though these cover 39\% of the field of view. Note that the source density in B\,59 turns into an underdensity without the counterparts of known YSOs. This comparison of X-ray source densities in the on-core and off-core regions shows that there is no indication for a significant, previously unknown population of PMS objects. 

In the on-core regions of the XMM-\textsl{Newton} observations, we therefore find an upper limit of 6 new evolved YSOs since this is the number of X-ray sources with appropriate $K_S$-band counterparts, not counting the known YSOs. However, the fact that the two regions with any detections in the on-core regions, B\,59 and the PMR, both show underdensities with respect to the surrounding background source densities (not taking into account the known YSOs in B\,59) suggests that a more realistic upper limit is closer to zero. A more stringent upper limit results when additionally requiring detections at 24\,$\mu$m, assuming that this selects $\sim80$\,\% of the class~III sources (\citealp{cie07}, see above). Apart from the known YSOs, only two X-ray sources with $K_S$-band detections in the on-core regions have 24\,$\mu$m counterparts, both of them in B\,59, suggesting an upper limit of only a handful of class~III sources. The combination of the three selection criteria (2MASS, X-ray flux, and MIPS detection) selects at least two thirds of any class~III population.

Finally, we can additionally give strict upper limits for any evolved YSO population in the three entire fields to account for any more distributed populations. For this purpose, we cannot treat the off-core source density as entirely unrelated to the cloud. While we have found that all sources in the B\,68 and ring fields can be explained by a combination of chance alignments and weak background sources, there are 23 selected X-ray sources in the B\,59 area when only 3.5 would be expected from chance alignments. Subtracting the expected number of chance alignments and the number of known YSOs among the selected sources (7), we derive an upper limit of $\sim13$ new candidate YSOs in the entire B\,59 XMM-\textsl{Newton} field. Note that among the selected sources, the field contains six detections at 24~$\mu$m apart from the known YSOs, again suggesting a more modest upper limit for any additional more evolved YSOs. These sources would be in addition to the upper limit of 15 previously known candidate YSOs in B\,59 \citep{bro07,for09}.

\subsection{Barnard 59}
\label{sec_b59}

In addition to the analysis in the previous section, which focused on finding candidate YSOs other than the ones already known from mid-infrared imaging, we now take a more detailed X-ray look at the sample of YSOs identified by \citet{bro07}, using \textsl{Spitzer}. It turns out that half of their sources have X-ray counterparts (Table~\ref{tab_brookeX}). Note that [BHB2007] 1 and 2 are far off-axis in the XMM-\textsl{Newton} data and detected as a relatively bright unresolved X-ray source. As corroborated by near-infrared spectroscopy (Covey et al. 2010, \textsl{subm.}), sources [BHB2007] 5 and 17 probably are background giants and remain undetected in X-rays. The sources [BHB2007] 3, 4, 8, 10, 12, 14, and 19 remain undetected in X-rays as well. While [BHB2007] 3, 4, 12, and 19 are located in chip gaps of the EPIC-PN detector, they are covered by the two MOS imaging detectors. While the sources may be hidden by higher foreground extinction, they could also be the lowest-mass members with the lowest X-ray luminosities. The fact that we detect one of the most deeply embedded sources, [BHB2007] 11, indicates that high extinction may not be the dominant reason for these non-detections.

As listed in Table~\ref{tab_brookeX}, nine sources have sufficient X-ray counts on the EPIC-pn chip to warrant a closer look at their X-ray spectra. Fits to these spectra were obtained using absorbed single-temperature APEC hot-plasma emission in \textsl{Sherpa} with a fixed setting of the metal abundances to 0.3 times the solar value. The spectra can be reasonably well fitted with such a model setup and the derived plasma temperatures are on the order of several 10$^7$~K. The derived absorbing column density can be empirically translated into magnitudes of visual extinction (see Section~\ref{sec_rosat}). The resulting values indicate extinction at levels consistently and considerably lower than the values derived from the extinction map (see Table~\ref{tab_brookeX}). However, while the extinction map traces the entire extinction along the line of sight toward the background sources, the X-ray extinction measures the extinction toward the respective source. This indicates that the detected sources are located toward the front of the cloud and the undetected sources may indeed be hidden by higher foreground extinction.
We have also extracted X-ray light curves for these sources. [BHB2007] 9, 13, and 16 -- all located in the inner cluster -- show conspicuous flares, likely of coronal origin, corroborating their classification as YSOs.

We note that two of the brightest X-ray sources in the B\,59 field, with more than 3000 counts each, are located in the off-core area. These are 2XMMi J171149.9--272050 and 2XMMi J171214.1--272140. Both have optical, near- and mid-infrared (IRAC) counterparts, and the latter source has a 24\,$\mu$m counterpart. Their SEDs look like stellar photospheres (unextincted for 2XMMi J171149.9 and $A_V\sim1.2$~mag in the case of 2XMMi J171214.1), even though the 24\,$\mu$m point of XMMi J171214.1 falls below a fit. Compared to extinction map estimates $A_V\sim4.7$~mag and $A_V\sim3.7$~mag, the low extinction as derived from the SEDs indicates that the sources are located in front of the cloud. The effective temperatures derived from the SED fits are on the order of 3500~K, i.e., these sources might be active M dwarfs. 2XMMi J171149.9 shows two flares in its X-ray lightcurve, indicative of coronal activity, while 2XMMi J171214.1 only shows slow variation. 
The X-ray spectra of both sources indicate very low levels of extinction, consistent with the SED results. The X-ray spectrum of 2XMMi J171214.1 can be well fitted by absorbed single-temperature hot-plasma emission while more than one component would be needed for 2XMMi J171149.9. Note that in both cases, also the X-ray to optical flux ratios would be consistent with an identification as M dwarfs\footnote{see, e.g., \citet{agu09} for stellar X-ray to optical flux ratios; with optical photometry from the Guide Star Catalog 2.3 \citep{las08}}.
A third bright X-ray source in the southern part of the B\,59 field, 2XMMi J171153.2--273607 does not have an optical or 2MASS counterpart and is only covered in two IRAC bands (1 and 3); it is also not covered at 24\,$\mu$m (note that it seems to have a counterpart in the two IRAC bands).

\section{Summary}
We have used X-ray data to probe the Pipe Nebula complex for signs of a population of class~III pre-main sequence stars. Such a population would have been missed by our previous work using mid-infrared excess emission to identify YSOs. To survey the entire complex, the only complete dataset is the relatively shallow ROSAT All-Sky Survey. Due to the nearby location of the Pipe Nebula, however, the absolute sensitivity of these data is still meaningful since we would expect about half of a population of G and K PMS stars to be detected. We find a deficit rather than an excess of ROSAT X-ray sources in the Pipe Nebula relative to its immediate surroundings. This finding is consistent with the X-ray population being dominated by background sources in the Galactic bulge, partly extinguished by the foreground Pipe Nebula. Only a single ROSAT source has an error circle that overlaps with one of the previously identified extinction cores, i.e., at most one ROSAT source is associated to the on-core regions.

In a more detailed look, we analyze three archival XMM-\textsl{Newton} pointings of prominent regions in the Pipe Nebula, encompassing B\,59, the PMR, and B\,68. We analyze the infrared SEDs of all X-ray sources with near-infrared counterparts ($K_S<12.4$~mag) in the on-core regions (plus most in the off-core regions, limited by data availability) and find that all sources in the on-core regions and all but one source in the off-core regions can be explained by extincted stellar photospheres. Among these three observations, the known cluster of YSOs in B\,59 is the only region where the density of X-ray sources in the on-core regions is indeed higher than the background source density. Without the known YSOs, B\,59 also shows an underdensity of sources in the on-core region, indicating that there is no significant YSO population beyond the sources that are already known. Roughly half of the previously known \textsl{Spitzer}-identified YSOs have X-ray counterparts with typical spectra; some show clear signs of variability. 

In the on-core regions of the XMM-\textsl{Newton} observations, we have found an upper limit of 6 evolved YSOs since this is the number of X-ray sources with $K_S$-band counterparts, not counting the known YSOs. However, the fact that the two regions with any detections in the on-core regions, B\,59 and the PMR, both show underdensities with respect to the surrounding background source densities (not taking into account the known YSOs in B\,59) suggests that a more realistic upper limit is closer to zero. A more stringent upper limit results when additionally requiring detections at 24\,$\mu$m. Only two sources in the on-core regions of B\,59 and only four more sources in the B\,59 field fulfill that criterion. We therefore conclude from the available X-ray data that there is no indication of an extended population of class III objects in the Pipe Nebula, corroborating the previously stated lack of star formation activity.

{\it Facilities:} \facility{XMM}, \facility{Spitzer}, \facility{ROSAT}, \facility{2MASS}

\acknowledgments{We thank Eric Feigelson and Leisa Townsley for informative discussions. This work is partly based on observations obtained with XMM-\textsl{Newton}, an ESA science mission with instruments and contributions directly funded by ESA Member States and the USA (NASA). This publication makes use of data products from the Two Micron All Sky Survey, which is a joint project of the University of Massachusetts and the Infrared Processing and Analysis Center/California Institute of Technology, funded by the National Aeronautics and Space Administration and the National Science Foundation. 
This research has made use of the SIMBAD database, operated at CDS, Strasbourg, France. This project is based, in part, on observations made with the \textsl{Spitzer} Space Telescope, which is operated by the Jet Propulsion Laboratory, California Institute of Technology, under a contract with NASA. Support for this work was provided by NASA through contract no. 1279166 issued by JPL/Caltech. B.P. acknowledges support by the Deutsche Akademie der Naturforscher Leopoldina (Halle, Germany) under grant BMBF-LPD 9901/8-170.}

\newpage

\begin{deluxetable}{lrrrr}
\tabletypesize{\scriptsize}
\tablecaption{Source selection for the three XMM-\textsl{Newton} pointings\label{tab_2xmmi_sel}}
\tablewidth{0pt}
\tablecolumns{5}
\tablehead{
\colhead{} & \colhead{B\,59} & \colhead{PMR}   & \colhead{B\,68} & \colhead{total}\\
}
\startdata
$A_V$(tot)\tablenotemark{a} [mag]           & $A_V=5.4\pm3.5$   &   $A_V=8.7\pm2.6$  & $A_V=1.2\pm1.2$  &     -- \\
no. 2MASS in FOV           &           10320   &             19594  &           13790  &  43704 \\
no. 2MASS ($K_S<12.4$~mag) &            1318   &              4625  &            1687  &   7630 \\
no. 2XMMi                  &             101   &                38  &             112  &    251 \\
no. 2XMMi w/$K_S<12.4$~mag &              23   &                 9  &              14\tablenotemark{b}  &     46 \\
no. 2XMMi w/$K_S<12.4$~mag on-core &      12   &                 1  &               0  &     13 \\
prev. known YSOs\tablenotemark{c}  &  $\leq15$ &                 0  &               0  & $\leq15$\\      
chance alignments\tablenotemark{d}&$3.5\pm1.9$ &       $6.6\pm2.6$  &     $6.9\pm2.6$  &     17.0\\
chance alignments on-core\tablenotemark{d}&$1.7\pm1.3$ &       $2.6\pm1.6$  &     $1.0\pm1.0$  &      5.3 \\
\enddata
\tablenotetext{a}{Mean $A_V$ and standard deviation in the three fields, from extinction map; see text.}
\tablenotetext{b}{We exclude nine low-extinction (and off-core) objects of these 14 sources; see text.}
\tablenotetext{c}{as listed by \citet{for09}}
\tablenotetext{d}{The chance alignments are the 2MASS sources ($K_S<12.4$~mag) that would randomly fall onto the combined area covered by the 3$\sigma$ 2XMMi search radii.}
\end{deluxetable}

\begin{deluxetable}{rrrrrrrrrl}
\tabletypesize{\scriptsize}
\tablecaption{2XMMi sources with 2MASS-$K_S$ counterparts ($K<12.4$~mag) in the on-core regions\label{tab_2xmmi}}
\tablewidth{0pt}
\tablecolumns{10}
\tablehead{
\colhead{No} & \colhead{Field} & \colhead{2XMMi} & \colhead{$A_V$(tot)\tablenotemark{a}} & \colhead{$A_V$(sub)\tablenotemark{a}}& \colhead{srcML\tablenotemark{b}} & \colhead{r(2MASS)\tablenotemark{c}} & \colhead{2MASS-$K_S$}& \colhead{err} & \colhead{ID}\\
\colhead{}   & \colhead{}      & \colhead{}      & \colhead{[mag]}      & \colhead{[mag]}     & \colhead{}      & \colhead{[$''$]}      & \colhead{[mag]}	    & \colhead{[mag]} & \colhead{}
}
\startdata
1  & B\,59 & J171059.9--272543 & 7.3 	 &  3.6   & 9.60E+01 & 2.70 & 10.30 & 0.02 &   \\               
2  & B\,59 & J171103.9--272256 & 7.9 	 &  4.2   & 2.86E+03 & 1.40 &  7.76 & 0.02 &  [BHB2007] 1/2  \\ 
3  & B\,59 & J171113.1--272936 & 7.2 	 &  2.6   & 1.13E+01 & 2.70 &  9.07 & 0.02 &   \\               
4  & B\,59 & J171116.4--272514 & 15.0    &  7.8   & 9.77E+03 & 1.41 &  8.75 & 0.03 &  [BHB2007] 6 \\    
5  & B\,59 & J171117.5--272535 & 13.0    &  8.5   & 2.29E+02 & 1.41 & 10.72 & 0.02 &   \\               
6  & B\,59 & J171121.5--272741 & 15.7    &  12.4  & 1.06E+03 & 0.55 &  8.98 & 0.03 &  [BHB2007] 9\\     
7  & B\,59 & J171127.0--272348 & 18.1    &  15.8  & 5.33E+03 & 0.10 &  9.08 & 0.03 &  [BHB2007] 13 \\   
8  & B\,59 & J171129.4--272536 & 15.6    &  12.4  & 6.19E+02 & 0.05 & 10.70 & 0.02 &  [BHB2007] 15 \\   
9  & B\,59 & J171130.3--272629 & 14.4    &  10.4  & 1.32E+04 & 0.26 &  8.89 & 0.02 &  [BHB2007] 16 \\   
10 & B\,59 & J171141.8--272547 & 8.7 	 &  4.5   & 9.48E+03 & 0.13 &  8.99 & 0.04 &  [BHB2007] 18\\    
11 & B\,59 & J171148.0--272633 & 8.5 	 &  4.9   & 3.70E+02 & 0.91 & 10.87 & 0.03 &   \\               
12 & B\,59 & J171231.5--271847 & 5.3	 &  2.5   & 1.76E+01 & 2.06 & 11.64 & 0.03 &   \\               
13 & Ring  & J173402.6--255219 & 10.2	 &  3.0   & 8.67E+00 & 1.88 & 12.21 & 0.03 &   \\               
\enddata
\tablenotetext{a}{visual extinction from extinction maps with the total extinction, $A_V$(tot), and with large-scale structure removed, $A_V$(sub) \citep{alv07}}
\tablenotetext{b}{2XMMi detection likelihood, see \citet{wat09}}
\tablenotetext{c}{distance to nearest 2MASS source}
\end{deluxetable}

\begin{deluxetable}{rrrrrrr}
\tabletypesize{\scriptsize}
\tablecaption{X-ray properties of YSOs in B\,59 \label{tab_brookeX}}
\tablewidth{0pt}
\tablecolumns{7}
\tablehead{
\colhead{[BHB2007]} & \colhead{total counts} & \colhead{$kT$} & \colhead{$N_H$}            & red.     & \colhead{$A_V$(X)\tablenotemark{a}} & \colhead{$A_V$(tot)\tablenotemark{a}} \\
\colhead{}       & \colhead{}     & \colhead{[keV]}& \colhead{10$^{22}$/cm$^2$} & $\chi^2$ & \colhead{[mag]}   & \colhead{[mag]}
}
\startdata
1\&2& 662\tablenotemark{b} &       &                        &      & 	      &      \\ 
3  &   nd &                        &                        &      & 	      &      \\
4  &   nd &                        &                        &      & 	      &      \\
5  &   nd &                        &                        &      & 	      &      \\
6  & 2032 & 1.95$^{+0.11}_{-0.11}$ & 0.63$^{+0.03}_{-0.03}$ & 0.82 & 3.2$\pm$0.2  & 15.0 \\ 
7  &  343 & 8.97$^{+8.16}_{-2.71}$ & 1.18$^{+0.16}_{-0.13}$ & 0.56 & 6.0$\pm$0.7  & 15.0 \\ 
8  & nd\tablenotemark{c}&          &                        &      & 	      &      \\ 
9  &  391 & 3.05$^{+0.35}_{-0.35}$ & 1.93$^{+0.18}_{-0.16}$ & 0.45 & 9.7$\pm$0.9  & 15.7 \\ 
10 &   nd &                        &                        &      & 	      &      \\
11 &   56\tablenotemark{d} &       &                        &      & 	      &      \\ 
12 &   nd &                        &                        &      & 	      &      \\
13 & 1289 & 1.64$^{+0.07}_{-0.07}$ & 0.67$^{+0.03}_{-0.03}$ & 0.88 & 3.4$\pm$0.2  & 18.1 \\ 
14 &   nd &                        &                        &      & 	      &      \\
15 &  285 & 0.98$^{+0.09}_{-0.09}$ & 0.86$^{+0.07}_{-0.06}$ & 0.40 & 4.3$\pm$0.3  & 15.6 \\ 
16 & 2700 & 1.95$^{+0.08}_{-0.08}$ & 0.53$^{+0.02}_{-0.02}$ & 0.62 & 2.7$\pm$0.1  & 14.4 \\ 
17 &   nd &                        &                        &      & 	      &      \\
18 & 2211 & 1.02$^{+0.02}_{-0.02}$ & 0.04$^{+0.01}_{-0.01}$ & 1.09 & 1.4$\pm$0.4  &  8.7 \\ 
19 &   nd &                        &                        &      & 	      &      \\
20 &  114\tablenotemark{d} &       &                        &      & 	      &      \\ 
\enddata
\tablenotetext{a}{optical extinction from the fit to the X-ray spectrum, $A_V$(X); and from the total-extinction map, $A_V$(tot).}
\tablenotetext{b}{blend, detected in EPIC-MOS only due to large off-axis angle; best fitted with two separate, extincted plasma components}
\tablenotetext{c}{possibly a weak X-ray source, but not picked up in source detection}
\tablenotetext{d}{poor fit; few data points}
\end{deluxetable}

\end{document}